# TRAINING NEW OPERATORS – THE FIRST SIX MONTHS

B. Worthel, FNAL, Batavia, IL 60510, USA

*Abstract*

The Fermilab Operations Department takes about two years to train a new Operator. The Operator's introductory (Concepts) On-the-Job-Training (OJT) gives him or her an overview of the laboratory, teaches the basic facts about all the accelerators, and it also teaches the new operator the training process used for all the rest of their OJT training. The Concepts OJT takes about four to six months for most people to complete. This paper will explain how this first six months of training sets the new employee on their path to becoming a fully trained Operator.

## FERMILAB OPERATIONS AND SHIFT WORK

The Operations Department is made up of 30 Operators (five crews of six people), 6 Crew Chiefs, 8 Operations Specialists, a Duty Assistant, an Associate Department Head, a Deputy Department Head, and a Department Head. That means that this group of 48 people (or less usually) is responsible for the operation of the accelerator complex.

Operators and Crew Chiefs work a rotating shift. New operators are repeatedly told and retold how much of a life change is needed to survive working shift and they all say that this is not a problem when they're hired, but when most people leave, they say it is because of shift work.

Operators work a rotating shift the majority of the time. The only exception is during long shutdowns. The rotation follows a five-week cycle, with three weekends off and two on.

## INITIAL TRAINING FOR NEW OPERATORS

During the first week, the new operator is inundated with training, training, and more training. Here are two lists: first is a list of safety training that's part of the New Employee Orientation classes and second is a list of safety training that occurs over the first four weeks. There is also training that is including in the OJT, but not included here.

*New Employee/User ES&H Orientation*
- Environmental Management System (EMS)
- GERT - (General Employee Radiation Training)
- Hazard Communication
- Introduction to Quality Assurance
- Job Hazard Analysis
- PPE (Personnel Protective Equipment)
- Sexual Harassment Training for FNAL Employees
- Traffic Safety Awareness

*Safety Training*
- Back Works Training
- Basic Computer Security
- CDF Supervised Access
- CPR
- Compressed Gas Cylinder Safety
- Computer Workstation Ergonomics
- Confined Spaces
- Cryogenic Safety (General)
- D0 Hazard Awareness
- Electrical Safety Orientation
- Electrical Safety in the Workplace (NFPA 70E)
- Fall Protection Orientation
- Fermilab Controlled Access
- Fire Extinguisher Use
- Hearing Conservation
- Lockout/Tagout Level 2
- Material Move Survey
- MuCool Test Area Hazard Awareness
- NuMI/MINOS Underground Safety Training
- O.D.H Training
- Oil Handling - Oil Pollution Prevention Program
- Pressure Safety Orientation
- Protecting Personal Information at Fermilab
- Radioactive Source Training (CR)
- Radiological Worker (CR)
- Scissors Lift Safety Training for CDF Personnel

When a new operator sees the safety training and then the amount of information that must be learned to operate the accelerator complex, it can be overwhelming. The On-the Job-Training method makes this learning possible by taking small steps at first and then building on that foundation.

The new operator receives the Concepts OJT and the Rookie Books on the first or second day after they've joined the department. They are encouraged to spend a lot of time with experienced operators in the Main Control Room watching how things work and meeting people.

After the first four weeks, or so, the new operator is assigned to a permanent crew and rotates with them through the normal schedule. The crew will be responsible for monitoring and carrying out training. Training is the top priority of the entire crew.

# THE ON THE JOB (OJT) METHOD OF TRAINING

The process works like this – each numbered training topic has a description of what is to be accomplished, along with a sign-off box to document this achievement. The sign-off box identifies the person who assures that the training is correct and complete. Many sign-off boxes identify this person simply as 'Trainer'. Experienced Operators (OP2s) most usually serve the role of 'Trainer', although any senior member of the Operations

```
┌─────────────────┐   ##.  Training Topic
│                 │   Description of what is to be accomplished.
│ Trainer   Date  │
└─────────────────┘
        ↑
   Trainers initials and date of completion entered here
```

Department may assume this task.

Here's an example of the format:

*Overview of the Training Process:*

This is a brief description of the training program that includes information on the role of a new operator, the trainer program, the training process, and the OP2 proficiency exams.

## 1. Roll of the new Operator

A new operator must take time to operate and tune each of the accelerators. At first, more experienced operators will guide the operator through the basics of turning on and off and tuning up each machine. Some aspects of tuning the accelerators are straightforward and can be learned in a short time. Other aspects are subtler and take a while to learn. Some aspects of tuning the accelerators change from week to week. There is no substitution for operational experience. In addition, each situation in the control room is unique, and different problems arise on a given day. New operators learn by applying themselves to a given situation, using concepts they were taught and common sense to solve a problem. Often problems cannot be fixed from the control room. To learn the basic troubleshooting skills, a new operator needs to visit the problem area accompanied with a more experienced operator. The problems encountered, will raise questions, and will give the new operator more opportunities to learn. This will not make the operator an expert, but it is essential in becoming a good operator. Participation is the best form of training (both in the Main Control Room and in the field), far better than any amount of reading material.

## 2. Training Materials

The most important training item is the "Introductory Training OJT." This is the first OJT an operator receives. It tracks an operator's progress by requiring signatures from trainers, other crewmembers, Crew Chiefs, and Specialists. The purpose of this OJT is to give an overview of the following items:

- The overall OJT training process
- The approximate training schedule
- An introduction to important personnel
- A guide to all of the required training and administrative processing
- An introduction to Main Control Room (MCR) operations
- An introduction to accelerator components and equipment
- An introduction to accelerator concepts
- The first OP2 Test.

### a. Rookie Books

There is a Rookie Book for each of the accelerators, including a book on basic accelerator concepts, a glossary of terms, and other miscellaneous material.

- Concepts
- Linac
- Booster
- Main Injector
- Tevatron
- Recycler
- Antiproton Source
- Controls
- Safety
- Glossary

These books contain background information on each of the machines, and are very helpful references. The systems for each of the accelerators often change as upgrades occur or operational conditions change, so occasionally material in the Rookie Books, or any training material for that matter, may become out of date. The crewmembers, Crew Chiefs, and Specialists help the new operator sort through what material is outdated and supplies the Training Specialist with updates.

### b. Additional Training Material

In addition to the Training Books, there are other study aids to help the newer operator become more accustomed to his/her job: training videos, self-tests, rookie books, and especially the other operators. The training committee is constantly expanding and updating many of these materials. The other operators and Crew Chiefs are the most important assets in learning the job. It wasn't too long ago that these crewmembers were new and they know how daunting an experience it can be. Becoming a good operator requires some work.

*Personnel Introduction*

The purpose of personnel introductions are to introduce the new operator to many of the personnel that they will interact with on a day-to-day basis in the while performing their jobs.

*Introductory MCR OJT*

The purpose of the Introductory Main Control Room (MCR) OJT is to start introducing the new operator to the procedures and tasks carried out by operators on a day-to-day basis in the Main Control Room. It also provides a guide to ensure that all administrative processing as well as lab wide, division, and safety training get administered in a timely fashion.

*Introductory Walkarounds OJT*

The purpose of the Introductory Walkarounds OJT is to start introducing the new operator to the locations and devices outside of the control room that will be important in the day-to-day operation of the accelerators.

*Introductory Accelerator Concepts*

The purpose of the Introductory Accelerator Concepts OJT is to introduce the new operator to the concepts, theory, and terminology that will enhance their understanding of the accelerators. There is an OP2 Introductory Test based on the information contained in this section and this section only.

*Introductory Self Tests*

The purpose of the Introductory Self Test is to allow operators to quiz themselves on the material that they are learning during their first 3 to 6 months. After all of the parts of this training package have been complete and the operator completes the self tests, and the operator has passed the Concepts Test based on material found in section 5, then the operator can continue on with the next stage of the training program.

## THE TRAINING OJT

The new operator must be aware that the OP2 trainer who signs off on the training is attesting that the new operator knows the information. Out-of-department system experts qualify as trainers on items under their jurisdiction. However, it is up to the new operator to make sure she or he really does know the information.

The basics of on-the-job training:
  a. The trainer will guide a new operator through an appropriate level of information for the sign off.
  b. A trainer will not necessarily sign off on a topic simply because she or he has explained the topic.
  c. In general, a trainer will expect the new operator to show that they know how to do something, find something, or explain some aspect of the accelerator complex before they will signing off on the topic.

So how do operators go about learning this information and getting a trainer's sign off?
  a. Watch what goes on in the MCR
  b. Participate in the MCR work and procedures
  c. Accompany experienced operators when they go out to fix problems
  d. Study the Rookie Books
  e. Ask crew members for help
  f. Talk to the Specialists and Crew Chiefs
  g. Talk to the people that were introduced in the Intro OJT
  h. Ask an out-of-department system expert
  i. Take good notes when talking with experts and use them to study
  j. Go ahead and have the trainer explain something again if it's unclear, just don't expect a sign off

After this fine definition of what an OJT sign off means, it must also be added that much of this is in the hands of the trainer. If she or he explains something to the new operator and thinks the operator knows the information and signs off on it, that is the trainer's option.

The Training OJTs were set up to expose each operator to the same minimal amount of material over a given period of time. It is extremely important to keep up to date in the program. Falling behind will only cause grief down the road. Here is a reminder that new operators are told over and over again — It is the ultimate responsibility of the operator to assure that his or her training is up-to-date and completed in a timely manner. Not only will this affect the new operator, but it will also affect the crew, Crew Chief, and the Operations Department.

A common question for operators in the training program is, to what level of detail should they learn the material? The best thing that can be said is to use some common sense. Going through the training material once to get signatures may or may not be adequate preparation for the OP2 walkaround or test. Certainly, having the System Design Engineer's knowledge of a system is not necessary either. A middle ground of understanding and the ability to apply this knowledge to specific operating problems is what's needed and expected.

What we have found is that an ideal Operator is a "jack-of-all-trades," having a good working knowledge of the accelerators, transport systems, as well as the control system. The Operators must have a basic understanding of many subjects, so during the first four months, as they go through the Introduction OJT, we impress upon the new Operators how important it is to not get bogged down in one area.

The whole crew is there to help the new operator train because it benefits them in the long run. Training is the operator's responsibility and part of their job! However, there are a lot of people willing to help them.

## SPECIALISTS

- Proton Source
- Pbar Source
- Main Injector
- Recycler
- Tevatron
- Controls
- Safety

## CONCEPTS EXAM

After the new Operator completes their OJT, they must take a OP2 written test. The Training Specialist creates a written exam for the OP1. OP1s are allowed ~2 hours to complete the written test, but it has never taken that long. The test is given in an area away from distractions (a quiet location).

The test format is, true or false, multiple choice, matching, labeling supplied drawings, and short answers. There are no essay questions.

A grade of 70% is required to receive a passing grade. However, if the unfortunate situation occurs and the operator does not obtain a score of 70% or greater, the operator will have to have a discussion with the Department Head. The Department Head will determine what steps will be necessary before rescheduling that written exam. In some cases, additional work may be assigned in the areas that were answered incorrectly. When a retake of the test is scheduled, the operator will have to achieve a score of 80% or greater for a passing grade.

Once the new operator has completed the Introductory OJT and passed the written OP2 Test, they receive the rest of the OJTs. Like the first OJT, these OJTs track the training progress by requiring signatures from crewmembers, other operators, Crew Chiefs, Specialists, or experts.

However, these OJTs are different from the Introductory OJT in their level of detail. Each Training document is broken into two sections: Main Control Room On-the-Job-Training (MCR OJT) and Walkaround OJT. The Main Control Room OJT will outline the procedures and tasks important to the day-to-day operations in the MCR. The Walkaround OJT will outline the devices and locations outside of the MCR that are important to the day-to-day operations of each of the accelerators.

## SUMMARY

The Fermi Operations Department has found through long experience that the OJT method of training works best for us. For people who like to jump in and start doing things, this style of training allows them to do it in a structured and safe way. For those who need to read about things first, we have lots of reference material, but we don't let them just study without doing; the OJT requires them to operate the systems.

**Acknowledgements**

The paper is based on information from the Fermilab Operations Introductory OJT, and from previously published work by Bob Mau, Dan Johnson, and Bruce Worthel.

Fermilab is managed by the Fermi Research Alliance under contract for the Unites States Department of Energy.

.